%
\input psfig.tex
\input aa.cmm
\font\small=cmr8
\overfullrule=0 pt
%
\MAINTITLE{The morphological segregation of galaxies in 
clusters.} 
\SUBTITLE{II. The properties of galaxies in the Coma cluster\FOOTNOTE{
Based on observations made with the 2-meter T\'elescope Bernard Lyot of
Pic-du-Midi Observatory, operated by INSU (CNRS), with the 3.6-meter
Canada-France-Hawaii telescope, operated by the National Research Council of
Canada, The Centre National de la recherche Scientifique of France and the
University of Hawaii, USA, with the Mayall 4-meter telescope at Kitt Peak
National Observatory, operated by AURA, and with the Schmidt telescope at 
the Calern Observatory (OCA)}}
\AUTHOR{Stefano Andreon}
\INSTITUTE{Istituto di Fisica Cosmica del CNR, via
	    Bassini 15, 20133 Milano, Italy\FOOTNOTE{Present adress}\newline
	    CNRS-UMR 5572, Observatoire Midi-Pyr\'en\'ees, 14, Av. E.
	    Belin, 31400 Toulouse, France\newline
	    andreon at ifctr.mi.cnr.it
	             }
\DATE{ Received January, 4, 1996; accepted April, 4, 1996} 
\ABSTRACT{
We have looked for differences in the galaxy properties along the Hubble
sequence and for the dependence of these properties on the environment, in
an absolute magnitude complete sample of 187 galaxies in the Coma cluster. 
The morphological type of all galaxies was determined from our own high
resolution data.  We also compared this sample with other published
complete samples of galaxies in the Perseus and Virgo clusters and in the
local field. 

Ellipticals and lenticulars are highly homogeneous in all their internal
photometric properties.  These galaxies are well described just by their
morphological type and luminosity, regardless of the environment. On the
other hand, some of the external properties of these galaxies and of their
subclasses differ markedly.  Then, for early-type galaxies, the
environment determines the space density of each type but not the internal
properties of the type, which are the same in all environments. 

Spirals form a heterogeneous class of objects whose photometric properties
depend on one extra parameter.  As a result, spirals that are blue in the
optical also have blue ultraviolet-optical colors, higher mean surface
brightness (or smaller radii) for their magnitude, strongly avoid the
cluster center and have high velocities relative to the cluster center.
The spatial distribution of spirals as a whole class is uniform.  For
spirals, therefore, the environment does not determine the space density,
which is the same in all environments, but strongly affects the internal
galaxy properties. 

The galaxies in Coma, as in Perseus, are segregated primarily with respect
to the supercluster main direction, thus providing a new interpretation of
the morphology-radius and morphology-density relations. 

Finally, the NGC 4839 group is richer in early-type galaxies than other
regions at the same distance from the cluster center, showing that the
morphological segregation could be a useful tool for discriminating
fortuitous alignments from real groups.} 
\KEYWORDS{ Galaxies: 
		elliptical and lenticular, cD, -- spiral --
		fundamental parameters  -- luminosity function, mass function
 		-- evolution -- interactions --
		cluster: individual: Coma cluster} 
\THESAURUS{03((11.03.4 Coma cluster;  11.05.1; 11.06.2; 11.05.2; 11.09.2;
11.12.2; 11.19.2)} 
\maketitle 
\titlea {Introduction}

The study of the morphological dependence of galaxy properties is one of the
major challenges of observational astronomy, since it can offer clues for 
understanding galaxy formation and evolution and is also central to many 
problems in cosmology (e.g. galaxy counts). 

It is common wisdom that ellipticals crowd the central region of clusters, 
whereas spirals populate the cluster outskirts and the field (e.g. Hubble \& 
Humason 1931, Dressler 1980b, Postman \& Geller 1984, Sanrom\`a \& Salvador-
Sol\'e 1990, Whitmore, Gilmore \& Jones 1993).  Galaxy optical (e.g. Hubble \& 
Humason 1931 and, for UBVRI colors, Holmberg 1958) and ultraviolet (2000 \AA, 
Donas, Milliard \& Laget 1995) 
colors depend on the morphological type and the same holds 
for the luminosity function (Sandage, Binggeli \& Tammann 1985), and many 
other galaxy properties (e.g. HI content).  Near infrared (JHK) colors of Es 
and S0s are similar  within the error bars, (Recillas-Cruz et al. 1990), as 
well as infrared colors of the Ss stages (sub-types) (Gavazzi \& Trinchieri 
1989) separately. 

However the real situation is complex.  One widely debated question concerns 
the fundamental parameter determining the galaxy type and therefore the 
observed morphological segregation.  Whitmore Gilmore \& Jones consider 
that it 
is the distance from the cluster center, Dressler and Postman \& Geller that 
it is the local density, while others (Sanrom\`a \& Salvador-Sol\'e 1990, 
Andreon 1994) state that the present data and analyses do not allow one to 
distinguish between the two possibilities. 

Several other important questions still call for an answer.

-- Does the environment determine the density of each morphological 
type (and therefore the relative spatial distribution without altering the 
galaxies' internal properties) or does it modify the morphological type of the 
galaxies in such a way as to induce the observed morphological segregation? 

-- Is the scatter in optical colors in each morphological class intrinsic and 
independent of the environment (Visvanathan \& Sandage 1977), is it related 
to the galaxy environment (Larson, Tinsley \& Caldwell 1980; using part of the 
data listed in Visvanathan \& Sandage 1977), or is it largely due to errors in 
the type assignment or to an incorrect determination of galaxy distances? 

-- Are the luminosity functions of boxy and disky ellipticals different 
(Bender et al. 1989), or not (Andreon 1994)? What is the relative spatial 
distribution of the two classes: are boxy ellipticals preferentially found in 
high density regions and disky ellipticals in low ones (Shioya \& Taniguchi 
1993, Caon \& Einasto 1995) or is their distribution density independent, as 
in the Perseus cluster (Andreon 1994)? 

In order to give some insight into these and other questions, we have been 
conducting an observing program of galaxies in different environments, 
including nearby (apparently) relaxed clusters (Coma and Perseus), nearby 
clusters with evidence of substructures (Abell 194; Bird 1994) or
not embedded in superclusters (Abell 496; Malumuth et al. 1992) and distant 
clusters (cl0939+4713) and regions in the field and in the Coma supercluster. 

Since we are studying the galaxy properties along the Hubble sequence and in 
different environments, we need a catalogue of galaxies formed using selection 
criteria that do not depend on morphological type and environment. 
Samples complete in absolute magnitude and volume are representative 
samples of galaxies. Obviously, with this choice, all galaxies fainter than 
the absolute magnitude limit are not represented at all, but we are interested only in 
bright ($M<M^*+3$) galaxies.  Our major concern is not related to rare low 
surface brightness galaxies missed in most galaxy catalogues, but is related 
to common, bright galaxies lost because not observed. In our catalogue too, 
low surface brightness galaxies are missing, but this is because they have 
fainter absolute magnitudes, and not because we failed to notice them. 

Since we are interested in morphology dependent properties, the morphological 
type is the crucial quantity to be estimated. The observations used to 
classify the galaxies have to be of good quality, enough to perform the 
morphological classification, possibly with the same rest frame resolution 
as that offered by the Hubble Space Telescope for distant clusters. This 
implies seeing conditions of about 1 arcsec, or better, for Coma galaxies.  
Furthermore, great attention must be taken in the type determination, in order 
to limit the scatter of galaxy properties inside each morphological 
class due just to incorrect type assignment.  The type assignment, 
therefore, has to be:
\newline -- independent of galaxy properties other than the ones indicated 
in the definition of types (based on the geometrical shape of isophotes and on 
the major/minor axis surface brightness profiles) and 
\newline -- reproducible, to guarantee the quality of the type assignment and 
to avoid personal biases. 

These prescriptions have been satisfied for the galaxies in the Perseus cluster
(whose properties have been presented in Andreon 1994) and in the Coma cluster, 
whose properties we present in this paper. 

\titlea{Why Coma?}

The Coma cluster is one of the richer and nearby clusters of galaxies. At 
first sight, the cluster looks relaxed and virialized in the optical and in 
the X-ray. It was in fact designated by Sarazin (1986) and Jones \& Forman 
(1984) as the prototype of the relaxed and virialized cluster.  

A careful analysis of the spatial distribution of galaxies (Fitchett \& 
Webster 1987, Mellier et al. 1988), good X-ray images (Briel, Henry \& B\"ohringer
1992,  White, Briel \& Henry 1993), and a large velocity sample 
(Biviano et al. 1995) revealed 
some substructures in this cluster.  Around NGC 4938, 40 arcmin SW from the 
cluster center, there is an excess of X-ray luminosity, of galaxies and also 
of galaxies with a low velocity relative to NGC 4938. Another substructure has 
been detected deep inside, in the core of the cluster, around one of the two 
dominant galaxies, NGC 4874, although the situation is complex, due to the 
fact that, for this structure, the velocity centroid, the spatial centroid and 
the dominant galaxy are offset with respect to one another. The presence of 
substructures in Coma has been interpreted as due to the collision of two 
cluster components, the main body of the Coma cluster and the NGC4839 group, 
seen just before (Biviano et al. 1995, Colless \& Dunn 1996) or just after 
(Briel, Henry \& B\"ohringer 1992) the first crossing encounter. 

The fact that such substructures are common in real clusters 
(Salvador-Sol\'e, Sanrom\`a \& Gonz\'alez-Casado 1993; Salvador-Sol\'e, 
Gonz\'alez-Casado \& Solanes 1993; Escalera et al. 1994), even in clusters 
previously considered virialized, again makes the Coma cluster interesting, 
since its galaxies lived the typical life of galaxies. 

For this reason, and taking into account the advantage that many properties of 
the galaxies in Coma are available in the literature, we focussed our 
attention on this cluster, studying in particular the variation of galaxy 
properties along the Hubble sequence from the center to the far outskirts of 
the cluster. 

\titlea {The sample and the physical parameters}

In order to avoid the biases arising in a poorly selected sample, we studied a 
complete flux limited sample formed by all galaxies brighter than B magnitude 
16.5 within one degree from the center of the Coma cluster (Figure 1, 
upper-left panel, 190 galaxies). At the distance of Coma, one degree 
corresponds to 2.7 Mpc ($H_0=50$ km s$^{-1}$ Mpc$^{-1}$). 

The following physical parameters were considered. \newline Positions, B 
isophotal magnitude, isophotal radius at the 26.5 B magnitude arcsec$^{-2}$, 
often called optical size in the literature, and the $B-R$ color were taken 
from Godwin, Metcalfe \& Peach (1977, hereafter GMP). Position, B magnitude 
and radius are known for all galaxies, and $B-R$ color for 92 \% of them. 

Galaxy radial velocities were compiled from a survey of literature data 
(Kent \& Gunn 1982, Caldwell et al. 1993, Biviano et al. 1995 and Colless \& 
Dunn 1996) and of public databases, such as Leda\fonote{Leda is the 
Lyon-Meudon Extragalactic Database supplied by the LEDA team at the 
CRAL-Observatoire de Lyon (France)} and NED\fonote{NED is the NASA/IPAC 
extragalactic database, operated for NASA by the Jet Propulsion Laboratory at 
Caltech.}. Radial velocities are known for 91 \% of the galaxies. For each    
galaxy, we adopt the median value of the published velocities.  Three galaxies of the 
sample (GMP2796, GMP4366, GMP4592) have a velocity larger than 11000 km/s, 
and, therefore, they do not belong to the Coma cluster.  We discard these 
three galaxies from our sample. 

Ultraviolet (2000 \AA) magnitudes were taken from Donas, Milliard \& Laget
(1995). Their UV 
flux-limited sample is a 90 \% complete sample down to UV magnitude 18. 
Because of the large scatter in galaxy UV-B colors (4 magnitudes), only a small 
number of galaxies (55) have a UV-B color, but the bias (selection in B) is well 
known and independent of the morphological type. 

Infrared ($J-H$, $H-K$) aperture colors were taken from Recillas-Cruz et al. 
(1990) who observed galaxies brighter than photographic magnitude 16, as 
estimated by Dressler (1980a), mostly within 46 arcmin from the center.  
By selecting galaxies brighter than B magnitude 16 within 46 armin of the center, 
we form a 83\% complete and unbiased subsample of 95 galaxies. 

The morphological types of the galaxies were taken from Andreon et al. (1996) 
and Andreon et al. (in preparation). These types supersede, or are as good as, 
the ones from the literature, because of the better resolution of the data 
used for the classification and because our type are based on the presence
of structural components (disk, bar, halo, etc.
see Andreon et al. 1996, and Andreon \& Davoust, in 
preparation, for details). For galaxies appearing more than once in these two 
papers, we adopt the type obtained from the better observational material 
used.  All but 3 blended galaxies are classified in one of the following 
detailed morphological classes: 
\newline boE, unE, diE, SA0, SAB0, SB0, S0/a, S, Irr\newline where 
boE and diE stand for boxy and disky ellipticals respectively, and the unE 
class is a blend of two classes: ellipticals that are not diE and not boE and 
ellipticals without evidence of boxiness or diskiness. The other types refer 
to the Hubble classes. As usual, E=boE+unE+diE, S0=SA0+SAB0+SB0 and early-type 
galaxies are Es+S0s.  boEs and unEs were merged in the (bo+un)Es class for 
physical (Bender et al. 1989) and statistical reasons, whereas the 
S0/a and the Irr classes were merged with the S class, in the spirit of the 
Hubble (1926) definition. S0/a are S by definition (see e.g. de Vaucouleurs
et al. 1991, p. 15). 

The panels of Fig. 1 show the spatial distribution of all the galaxies, for 
each morphological type, and of the 3 galaxies without detailed type 
(blended galaxies). 

The following quantities are computed for further analysis. \newline
-- GMP's x, y positions have been rotated clockwise by 45 degrees (around
the cluster center adopted by GMP, $\alpha=12^h57^m18^s$, $\delta=+28^o14'24"$ 
(1950.0)) to align the positive x semi-axis with the direction of the
NGC4839 group and with the main direction of the Coma/A1367 supercluster; 
\newline
-- the clustercentric distance, the distance to the nearest neighboring galaxy 
($d_1$) and the local density ($\rho_{10}$), defined as the density inside the 
smallest circle containing the 10 nearest galaxies (Dressler 1980a) have been 
computed. These three quantities are obviously not independent. Figure 2 
shows the tight relation between the clustercentric distance and the local 
density. The spread at $r\sim40$ arcmin is due to the NGC4839 substructure. 
\newline 
-- the mean surface brightness, $<$SuBr$>$, has been computed taking into 
account the galaxy ellipticity. 

\titlea {The method}

The method used to test the reality of the differences (if any) between the 
galaxy properties along the Hubble sequence has been presented in Andreon 
(1994).  In summary, the probability that an observed difference is real is 
computed by Monte Carlo simulations in which the types of the galaxies are 
shuffled randomly. We did $10^5$ simulations for each pair of classes and for 
each galaxy property analysed. We considered the following statistical 
quantities:  the {\it t} value of the T test, the {\it f} value of the F test, 
the tail and skewness indices ti and si, that measure differences between the 
mean, the dispersion, the curtosis and the skewness, respectively, of pairs of 
distributions. We used also the D statistics of the Kolmogorov-Smirnov test, 
and the P5 probability of the vector (t, f, si, ti, D). In the determination 
of the f and t values of the F and T tests, we replaced the mean and the 
dispersion with equivalent robust quantities, that are less sensitive to 
outliers.  These are the central location index and the scale index 
(Beers, Flynn \& Gebhardt 1990) which, for the sake of clarity, we continue to 
call mean and dispersion. We verified that our results are insensitive to the 
choice of the exact form for the mean and dispersion, on the condition that 
they are robust. 

We tested the agreement between the probabilities computed using Monte Carlo
simulations and the theoretical ones for the two tests (K-S and classical T
tests) for which the theoretical probability function is known, {\it
whatever} the assumed parental distribution from which the data are
taken. The comparison concerns real probabilities measured during the
statistical analysis and not uniform deviates, as in Andreon (1994).
We found an excellent agreement for the T and K-S tests 
over four decades in probability. 

\titlea {Results and comparisons with previous studies}

Two degrees of resolution were adopted for the morphological type
during the comparison of the galaxy properties: a coarse one (E, S0, S),
typical of many works on the morphological segregation of galaxies in
clusters, and a finer one ((un+bo)E, diE, SA0, SB0, S).

Fig. 3 summarizes our examination of the morphological dependence of the 
galaxy properties. Some differences are clearly outstanding, whereas some 
others are difficult to establish. Table 1 lists the measured probabilities 
when we found differences larger than ``$3\sigma$" (or, better, when 
the probability that the difference is due to random fluctuations is less than 
0.002). These graphical and tabular data are discussed in the following 
sections where we present first our results, then those of the 
literature, and discuss both together. Null results are commented only when 
interesting. 

\titleb {B Luminosity}

Statistically, (un+bo)Es are brighter than all other types, except diEs, which 
have a similar luminosity function. The bulk of SA0s are fainter than the 
typical diEs, which have a flatter luminosity distribution.  The SA0 
luminosity function is marginally narrower than the S one. As a consequence, 
Es are brighter, on the average, than S0s and Ss.  

The shapes of the bright part ($M_B<-19.2$) of the luminosity function of 
Hubble types are similar to the ones of the field and of the Virgo cluster 
(Binggeli, Sandage \& Tammann 1988) and to the one computed by Thompson \& 
Gregory (1980) for a sample of galaxies of Coma largely overlapping our own, 
but with fully independent magnitudes (from Godwin \& Peach 1977) and (their 
unpublished) morphologies:  the S0 and S luminosity functions increase more 
sharply than the E one, starting from fainter luminosities. On the contrary, 
the morphological compositions (i.e. the normalizations of the luminosity 
function) of the Coma and Virgo clusters differ. 

These results support the finding of Binggeli, Sandage \& Tammann (1988) that 
the variation of the bright part of the global luminosity function is related 
to the variation of the morphological composition and that the luminosity 
function of each type is universal. However, one exception is known: in the 
Perseus cluster (morphologies taken from Poulain, Nieto \& Davoust
1991 and magnitudes 
taken from Bucknell, Godwin \& Peach 1979) the S0 luminosity function does not 
differ from the E one (Andreon 1994), as it starts at the same luminosity and 
flattens off at $M_B \sim -20.5$. Misclassifications between Es and S0s cannot 
explain the anomaly of Perseus' S0s, since the rest-frame resolution of the 
data used to classify the galaxies in Perseus is {\it better} than the one 
used for Coma. 

The luminosity functions of boEs and diEs of Coma galaxies are compatible, 
within the large errors due to small statistics, with the ones for Perseus 
given by Andreon (1994), but both are shifted faintward by one magnitude with 
respect to the preliminary ones obtained by the ESO-Key Program team ``Toward 
a physical classification or early-type galaxies" (Bender et al. 1993) for a 
larger and more heterogeneous sample.  Photometric (systematic) errors account 
for a shift of only 0.11 magnitude (necessary to translate our B isophotal 
magnitude into the asymptotic B magnitude used in the comparison sample).  
Errors in the morphological classification between the two E classes cannot 
explain this result, since both classes are affected by the same magnitude 
shift.  
This shift must be due to a selection bias, otherwise it would imply that Es, 
as a whole class, do not have the same luminosity function in different 
environments, which is in contradiction with observational evidence. 
It is possible that Bender et al. oversampled the bright end of the boE and 
diE luminosity functions and that galaxies of normal luminosity are missing, 
thus explaining the observed result, since their sample is flux limited  
at a bright apparent magnitude ($B=13$) but not absolute magnitude complete. 

\titleb {Optical Size ($r_{iso}$)}

(un+bo)Es are larger than all other types but diEs. diEs are a bit larger and 
their distribution is flatter than later types. Consequently Es are larger 
than S0s and Ss. The same applies for the coarse classes, Es {\it vs} S0s and 
Ss. However, since Es are brighter and brighter galaxies also have more of 
everything, our results are dominated by differences in the luminosity 
functions. We stress therefore that this result, as well as the similar one 
obtained by Roberts \& Haynes (1994), just reflect the luminosity and 
morphological composition of catalogued galaxies and are not general 
properties of the Hubble types. All the other quantities presented in the 
following are normalized to the galaxy luminosity, and, therefore, the above 
caveat does not apply. 

\titleb {Mean Surface Brightness ($<$SuBr$>$)}

(un+bo)Es have a higher mean surface brightness than all the other classes but 
diEs. The same applies to the E class compared to S0s and Ss. Therefore, 
at fixed magnitude, early-type galaxies are larger than other types. Such a 
difference was not found by Roberts \& Haynes (1994) in their sample, either 
because Coma galaxies are peculiar or because Roberts \& Haynes missed early-
type galaxies of mean surface brightness fainter than the typical one of 
spirals. 

Es in Coma are probably not peculiar, since they have the correct dimension 
for their magnitude, and there is no difference in the fundamental plane for 
Coma and non-Coma galaxies (Bower, Lucey \& Ellis 1992, Guzm\'an et al. 1992, 
J\o rgensen, Franx \& Kjaergaard 1993, J\o rgensen, Franx \& Kjaergaard 1996). 

The dispersions of the E and S0 mean surface brightness distributions are
compatible with the ones expected from the 0.13 mag error in the galaxy
magnitudes estimated by GMP, and a 13\% error in the size measurement,
which is a reasonable estimate of that quantity. The width of the S mean
surface brightness distribution is larger and must be due to correlations 
between the mean surface brightness and other galaxy properties, as shown in 
Sect. 5.10. 

\titleb {$B-R$ Color}

The color distribution of spiral galaxies markedly differs (as expected) from 
those of the other coarse and detailed types, because there are no blue ($B-
R<1.6$) early-type galaxies. Es and S0s have the same color (1.87), and no 
dispersion is present in the two color distributions besides that due to 
photometric errors.  The presence of a bar does not change the $B-R$ color of 
lenticulars at a statistically significant level. The same applies for the 
presence of a disk in early-type galaxies. The (un+bo)E and SB0 distributions 
have opposite skewnesses, but we must be cautious in this case, because we are 
comparing the shape of two distributions blurred by photometric color errors. 
Ss have a color distribution which is incompatible with the hypothesis that 
all Ss have the same $B-R$ color, and, as a result, the spiral color is 
correlated with the $<$SuBr$>$ and the galaxy location in the Coma cluster, as 
shown below. 
 
\titleb {UV-B Color}

Because of the low number of galaxies detected both in B and UV, we only 
compare the coarse classes.  The median colors are 2.92, 2.22, -0.28 
for the E, S0 and S classes, respectively, for our B selected sample. We give 
statistical significance to the finding of Donas, Milliard \& Laget (1995) that S galaxies 
have bluer UV-B color than early-type galaxies (E, S0 and E+S0). The E and S0 
color distributions show a dispersion of the same order as the photometric 
accuracy, whereas the S one has a dispersion twice as large. $UV-B$ is 
correlated to $B-R$, bluer galaxies being bluer in both colors. The nature 
of this dispersion is discussed below, together with all the other already 
mentioned ones. 

\titleb {$J-H$ \& $H-K$ Infrared Colors}

There is no difference between the $J-H$ colors of the detailed and coarse 
morphological types, showing that the two filters map fundamentally the same 
luminosity emission. In $H-K$, Es are 0.04 mag bluer than S0s, and 0.09 mag 
bluer than Ss. A dispersion of 0.07 mag was measured for the E and S0 
color distributions, which can be almost completely accounted for by 
photometric errors, whereas a larger spread is found for Ss (0.12 mag). 
In $H-K$, (bo+un)Es are bluer than SB0s and Ss, the latter showing also 
a large range in color. 

Infrared colors are not expected to depend on the morphological type in the 
hypothesis that all emission is due to stars. In fact, in the JHK bands, the 
shapes of the spectra of galaxies with very different star formation 
histories look very similar, except during the first million years for an 
instantaneous burst (Bruzual \& Charlot 1993, their Fig 4). Therefore we must 
pay attention to other effects possibly mimicking the detected correlation.  
Visual inspection of the infrared color magnitude diagrams for each 
morphological type reveals that the reddest galaxies are Ss, independently of 
the magnitude, excluding the possibility of a magnitude-color effect.  

We looked for confirmation of our finding in the literature for magnitude 
complete samples. Recillas-Cruz et al. (1990), from which we obtained the 
infrared colors, did not look for differences in colors between early-type 
galaxies and Ss. Gavazzi \& Trinchieri (1989) observed only spiral galaxies 
in the Coma supercluster and they found constant infrared colors from Sa to 
Irr.  The situation is actually even more complicated, since the $H-K$ median 
colors of Gavazzi \& Trinchieri's Coma supercluster spirals are similar to 
those of our Es and, therefore, redder than those of our Ss. Some caution is 
necessary: Gavazzi \& Trinchieri's and Recillas-Cruz et al.'s spirals are not 
measured within the same aperture, Gavazzi \& Trinchieri's sample is not 
truly complete (and therefore could be biased) and an environmental effect 
could be present, virtually all of Gavazzi \& Trinchieri's galaxies being 
more than 3 Mpc away from the cluster center and all of ours inside that 
distance. On the other hand, five galaxies in common between the two samples 
have equal $H-K$ colors within $\sim$ 0.03 mag and in $J-H$ no differences are 
found between the two spiral samples. Photometry of a large, complete and more 
uniform sample of Ss in different environments, through an aperture large 
enough to encompass any eventual central non stellar emission, is necessary to 
solve this puzzle. 

\titleb {Morphological segregations}

The diE x-distribution is more centrally concentrated than all the other ones, 
which are compatible with one another (see also Fig 1).  The SB0 
y-distribution is more centrally concentrated than all but the (un+bo)E one.  
As a result of these two segregations, Es and S0s are more centrally 
concentrated in the x and the y directions, respectively, than Ss. 

Spirals are uniformly distributed over the studied
field (see fig. 1), where the galaxy density varies over four
decades, the distance from the cluster center varies from 0 to 2.7 Mpc
and the distance to the nearest galaxy from 0 to 600 Kpc. 

The nearest galaxy distance of Ss differs from those of the coarse E
class and of its two detailed classes. This is in agreement with the
findings of studies comparing the correlation function of Ss and Es (e.g.
Davis \& Geller 1976). 

The only, although marginal radial
segregation present, concerns diEs, that are spatially more centrally 
concentrated than Ss. 

No statistically significant radial or density segregation
is detected; the same result was found for the Perseus cluster (Andreon 1993). 
As a check, we verified that no statistically significant radial segregation 
is found for Coma galaxies even using Dressler (1980a) morphological types 
instead of our own, and that our computed densities agree with Dressler's, 
showing that the lack of detection of the morphology -- radius and morphology-
density relations is not due to the possibility that our types or densities, 
respectively, are wrong. 

As stressed in Andreon (1993), the presence
of a segregation with respect to a privileged direction
naturally implies a radial or density 
segregation, the primary cause of the segregation is the 
former and not the latter, because one is 
statistically significant and not the other.

Lenticulars are overabundant at all radii and this make 
the morphology -- radius relation not 
compatible with the standard one presented in Whitmore, Gilmore \&
Jones (1993).

These two facts suggest that, even if there is a general tendency for galaxies 
to obey morphology -- density (or radius) relations, evidenced by averaging 
them over many clusters, as done by almost all previous studies (see 
references in the introduction), there is a large scatter in the individual 
clusters which can affect the morphology-density (or radius) shapes or 
normalization, as in Coma and in Perseus. 

We failed to find a radius or density segregation in the present 
sample of Coma galaxies, but we did find strong evidence 
for a different location of the spatial distribution of the types with respect 
to a privileged direction, the main direction of the supercluster in which 
they are enbedded. The same failure to detect a radius or density morphological 
segregation and the successful detection of a privileged direction in the 
cluster also occurred in the case of Perseus, although we did not recognize at 
the time that the preferred direction was the main supercluster's direction. 
Many types are concerned, and not only spirals, as expected if the infalling 
galaxies were mainly spirals and if most of the infall occurred along the 
supercluster's main direction.  Incidentally, the detection of a privileged 
direction for the spatial distributions of Ss is very difficult from a 
statistical point of view, because their spatial distribution is fairly 
uniform over the observed fields. It is normal, therefore, that their spatial 
distribution looks almost circular in Figure 1. 
 
We can exclude that the preferred direction that we found is due to biases in 
the morphological classification, for two reasons. First of all, in our scheme, 
type assignment does not depend on personal bias; in other words, it is 
reproducible at more than 95\% (Andreon \& Davoust, in preparation).  Secondly, no 
personal influence of the author on the type assignment is possible because 
the type assignment was done before we understood that the preferred 
direction is the supercluster one, either by people not in contact with 
author at 
the time or by two morphologists independently, often not including the 
author. 

We have also verified that the quality of the observations does not depend 
on the found preferred direction.

\titleb {Morphology-velocity segregation}

Es and, marginally, S0s have a smaller velocity dispersion than Ss (758, 695 
and 1325 km s$^{-1}$ respectively). The same happens for (un+bo)Es vs Ss and, 
marginally, for SA0s vs Ss. Blue ($B-R < 1.7$) spirals have a velocity 
dispersion of 1600 km s$^{-1}$ whereas red ones have a dispersion of 940 km 
s$^{-1}$.  The SB0 velocity distribution differs from the (un+bo)E one and, 
marginally, from that of SA0s, mainly because of a small velocity offset (540 
and 380 km s$^{-1}$ respectively). The qualitative behavior we found for the 
velocity distribution of the coarse Hubble types is identical to that of the 
smaller data set of Zabludoff \& Franx (1993), but their $2\sigma$ detection 
of different means is not confirmed using our larger sample and our robust 
measure of the mean. The spiral velocity distribution is also similar to the 
ones of Gavazzi (1987) and Donas, Milliard \& Laget (1995). 

Simple arguments show that the ratio of the velocity dispersion of infalling 
galaxies to virialized galaxies is $\sqrt{2}$ for an isolated and spherical 
cluster. This prediction is verified in Virgo (Huchra 1985) assuming that the
infalling galaxies are Ss and that the virialized ones are Es and S0s, and in 
Coma (Colless \& Dunn 1996) using the color-magnitude diagram to split the 
sample into infalling and virialized galaxies. Using our morphological types, 
we found that $\sigma_{S} = \sqrt{3} \times \sigma_{E}$ and $\sigma_{blue S} = 
\sqrt{4} \times \sigma_{E}$. 

This means either that the hypothesis of a spherical isolated infall is not 
verified in Coma, or that infalling galaxies are not only Ss or blue ($B-
R<1.7$) Ss, but even Es and S0s. 

\titleb {The NGC 4839 substructure}

We removed from the sample the 11 galaxies whose distance to the NGC 4839 
group is less than $\sim400$ Kpc, in order to diminish the effect of this 
substructure on the galaxy properties.  Only negligible changes appear, all of 
them concern comparisons involving SA0s, which compose the bulk of the flagged 
galaxies. 

If the morphological segregation is universal, then a substructure (a local 
density enhancement) in a cluster can be detected as a local change of the 
morphological composition associated with a local density enhancement. Then, 
the morphological segregation, as the X-ray map, could allow one to 
disentangle fortuitous alignments (which do not change the local morphological 
composition) from real density enhancements. We have tested this possibility 
for the NGC 4839 group. The morphological composition of the clump 
(E/S0/S=20/60/20) is richer in S0s and poorer in Ss than the mean composition 
of the cluster at the same radius (E/S0/S=21/33/45), giving a positive result 
to our test. 

\titleb{Color dependent segregations} 

Up to now, we have compared the properties of the galaxies of different types.  
Since galaxies of different colors in many distant and nearby clusters appear 
to populate different regions of the cluster (Butcher \& Oemler 1984, 
Schneider, Dressler \& Gunn 1986) and to have different properties (Whitmore 
1984, Staveley-Smith \& Davies 1988), we attempt a comparison of galaxy 
properties dividing our sample in color bins. 

Our statistical tests and inspection of the spatial distribution of red ($B-R > 
1.7$) and blue ($B-R < 1.7$) galaxies (Figure 4) confirm that the color 
segregation found for many clusters is present in Coma, too. Note that all but 
one blue galaxy are Ss and that the segregation is present in galaxy density 
even between red and blue spirals, in spite of smaller statistics. Adding to 
our sample all blue ($B-R < 1.7$) galaxies brighter than B magnitude 16.5, 
having a distance from the cluster center 
between 1$\degr$ and 1$\degr$23', the largest radius contained in GMP's region, 
confirms the impression that the highest density of blue galaxies is centered 
at 25 arcmin (1.1 Mpc) from the cluster center  and that the density of blue 
galaxies falls inside and outside this radius. In our sample, $UV-B$ is well 
correlated to $B-R$, and therefore the bluest galaxies in the optical are also 
the bluest ones in the UV. It is therefore normal that Donas, Milliard \& Laget
(1995) 
detect the largest UV flux density at this radius. We have infrared colors of 
too few blue galaxies (simply because too few of them have been observed) to 
say anything about the difference between infrared colors of red and blue 
spirals, although a larger spread than the observational errors was found in 
the infrared color distribution of spiral galaxies. 
 
The spectra of five of these blue spirals have been observed by Bothun \& 
Dressler (1986). They have  ``anomalously strong Balmer absorption lines" 
arising ``from a relatively recent star formation event" and the central 
surface brightness of all three of them (observed in imaging mode) is 
overbright. Since the UV is sensitive to recent star formation (see e.g. fig 4 
of Bruzual \& Charlot 1993), probably all our blue spirals have experienced  
an accelerated star formation rate. This is supported by the fact that, in 
Coma, all blue spiral galaxies, as Bothun \& Dressler's (1986) starburst 
galaxies, have a higher surface brightness (or are smaller) for their 
magnitude. In fact blue Ss have a higher mean surface brightness than 23.85 
mag arcsec$^{-2}$ whereas $\sim 50$\% of red spiral galaxies have a fainter 
mean surface brightness (Fig. 5).  Figure 5 also shows that red face-on 
spirals have a higher mean surface brightness than red edge-on galaxies.  
Geometrical or internal absorption effects cannot account for this behavior, 
since the observed luminosity of face-on spiral galaxies is fainter, not 
brighter, than edge-on spirals (see e.g. Boselli \& Gavazzi 1994). 

In principle, many physical mechanisms can accelerate star formation, as 
interactions with neighboring galaxies (e.g. Lonsdale, Persson \& Matthews 
1984) and ram-pressure (e.g. Gunn \& Gott 1972). In the first case, blue 
and red 
spirals are expected to have different nearest neighbor distributions. Our 
statistical test fails to reveal such a difference. 
Even more, we found that bluer spirals have high velocities relative 
to the cluster center and do not live in low velocity dispersion environments, 
as expected in this scenario (Bothun \& Dressler 1986). In the second case, 
blue spirals are expected to have higher relative velocities and to be 
segregated in the intercluster gas-density dimension.  Our statistical tests 
allow us to detect the first prediction and also to detect a galaxy density 
segregation (bluer spirals avoid the high densities of the cluster center) 
that it is an intercluster gas density segregation since gas and galaxy 
densities are clearly correlated in clusters. 

\titlea {The spiral fraction in the Coma cluster} 

The high spiral fraction we found in the inner region of the Perseus
cluster (Andreon 1994), a factor 3 higher than previous determinations,
and the systematic underestimate of the spiral fraction in nearby clusters
suggested by the redshift dependence of the observed spiral fraction
(Andreon 1993), prompted us to compare the spiral fraction of the Coma
cluster computed by Butcher \& Oemler (1984, hereafter BO) with ours.
This comparison is important, since BO's value of the spiral fraction is the
reference for the comparison with distant ($z\sim0.4$) clusters. The
importance of this point is enhanced by the fact that, after the refurbishing 
of the Hubble Space Telescope, we are now observing distant clusters with 
better rest-frame resolution than nearby clusters.  

Within 22 arcmin from the cluster center, which is the radius within which BO 
prescribe that the spiral fraction be computed for the Coma cluster, BO found 
77 galaxies brighter than B magnitude 16.5. Five of them were classified S by 
BO and by us, but 14 were classified S0 by BO and S by us. Half of these 14 
galaxies were classified S by Dressler (1980a) with an observational material 
of slightly lower quality than ours, confirming that the estimated spiral 
fraction of Coma is a lower estimate. 

Considering that the median seeing for our data is 1.2 arcsec, whereas an 
equal rest-frame comparison between Coma and distant ($z\sim0.4$) clusters 
requires a 1 arcsec resolution for the whole sample, the spiral
fraction in Coma is underestimated by a factor 3, or more, as in Perseus, 
confirming  the tendency of morphologists to underestimate the spiral 
fraction in nearby clusters, probably because of the low resolution data used. 

The spiral excess of galaxies in distant clusters with respect to nearby ones 
is therefore strongly reduced, although it seems that spiral galaxies in 
distant clusters do not look like nearby spirals (Dressler et al. 1994), 
and therefore that a 
Butcher-Oemler effect, in another form, is still present. 

\titlea {The environment and the properties of galaxies}

In principle, the environment has two possible effects on galaxy properties. It 
can limit itself to affecting only the space density of each morphological type.
This implies that all galaxies of the same type have the same properties, but 
different spatial distributions. Or it can affect the internal properties of 
galaxies (colors, mean surface brightness, shape of the luminosity function, 
etc.). These two possibilities can be discriminated, since in the first case 
the internal properties of galaxies are universal, whereas in the second case 
they depend on the environment. 

The results discussed in the previous sections, together with other 
observational evidence, in particular our results on Perseus (Andreon 1994), 
allow us to do such a discrimination. We limit ourselves to present epoch 
galaxies, leaving the discussion on the variation of the properties with epoch 
to a future paper.  

In the next paragraphs we also show that the coarse morphological appearance of the 
galaxies (i.e the E, S0, S Hubble types) splits well the galaxies of different 
physical types in different classes and brings together similar galaxies of 
early-type (i.e. Es and S0s), whereas the S class is a blend of galaxies 
with different properties.
Contrary to the case of coarse Hubble types, the detailed morphological 
appearance of Es and S0s (i.e. the boxiness or diskiness for Es or the 
presence of a bar for S0s) is not a successful indicator of different 
internal physical (considered) properties, but it traces well different 
external properties of the types. 

\titleb {Es \& S0}

Early-type galaxies (i.e. not spirals) look {\it homogeneous} in their
internal properties: no difference was found between the properties
of galaxies of same coarse morphological type, but of different detailed
type.  The considered properties are: luminosity function, mean surface
brightness, optical and infrared colors. The observed scatter in the
properties is the one produced by observational errors. Differences
were found between some properties of Es and S0s (the mean surface
brightness and the luminosity function). These two facts show that the
morphological type is a good way of splitting galaxies of different properties
in two classes and of bringing together similar galaxies.  The effectiveness 
of the Hubble morphological scheme for early-type galaxies is a well known 
fact, for example from the kinematical point of view. 

The luminosity function of Es and S0s seems universal: our results for Coma 
galaxies are fully compatible with the ones for the Virgo cluster and for the 
field, and, concerning ellipticals, with the ones for the Perseus cluster. The 
properties of early-type galaxies are not strongly dependent on the distance 
from the cluster center or the density, otherwise we would detect an 
inhomogeneity in the properties of these galaxies, and in environments as 
different as the rich Coma and Perseus clusters, the poor Virgo cluster and 
the local field. The color-magnitude relation for early type galaxies seems to 
be universal, independent of cluster richness, X-ray luminosity and redshift 
up to $z\sim0.2$ (Garilli et al. 1996).  The spectral appearance, and 
not just the photometric properties, are normal for early-type galaxies in 
Coma:  among the 57 (out of 127) early-type galaxies in our sample selected 
for spectroscopic study by Caldwell et al. (1993), only a negligible minority 
(4, i.e. $\sim 7\%$) have an abnormal spectrum (indicative of a starburst). 
Only slight, if any, systematic departures of the structure and dynamics from 
the homology have been found for Es when studying the fundamental plane 
(Pahre, Djorgovski \& de Carvalho 1996), perhaps depending on environment 
(Guzm\'an et al. 1993) or not (J\o rgensen, Franx \& Kjaergaard 1993). 

All this evidence suggests that Es and S0s are classes of galaxies with 
homogeneous properties and that the {\it internal} properties of early-type 
galaxies are not significantly affected by the interactions with the galaxy 
environment, with perhaps one exception (S0s in Perseus).

However, there is evidence that {\it external} properties of early -- type 
galaxies are affected by the environment.  The general tendency for galaxies 
to obey the morphology -- density or morphology-radius relations on the one hand, 
and the fact that both sub-classes of Es and S0s show differences in their 
relative spatial distributions in Coma and in Perseus and that their spatial 
distributions present preferred directions related to the supercluster's 
direction on the other hand, both point toward a relation between the presence 
of a galaxy in a determined environment and its morphological type. These two 
facts are not mutually contradictory: the {\it space density} of each morphological 
type in a given environment depends on the environment, but {\it not the 
properties} of the morphological types, that depend just on the morphological 
type and that are the same in all (studied) environments. 


\titleb {Ss}

The situation for spirals looks different. First of all, spiral galaxies
are {\it heterogeneous} in their internal properties: their mean surface
brightness and their ultraviolet, optical and infrared colors show a
scatter not accounted for by observational errors.  Moreover, the optical and
ultraviolet colors are correlated with each other and with the mean surface
brightness. Bluer spirals are not a random subsample of spirals, in the sense
that they have a higher velocity relative to the cluster center, 
they avoid the high densities of the cluster center and they are overbright
for their magnitude.

Therefore, the S class does not regroup similar galaxies, but it is rather a 
blend of classes of galaxies with different properties. The scatter in the 
internal properties is intrinsic, at all spiral stages as even incomplete 
samples show (for optical color, see Gavazzi \& Trinchieri 1989; for mean 
surface brightness, see Roberts \& Haynes 1994; for many properties of Sb and 
Sc galaxies, see Stavely-Smith \& Davies 1987 and 1988) and therefore the 
scatter we found in the coarse S class does not come from our coarse sampling 
of the spiral stages. This conclusion is reinforced by the fact that the 
variation of the properties along the spiral stages is small or absent (e.g. 
Gavazzi \& Trinchieri 1989 for infrared colors). 

Once we have established that the observed spread in the galaxy properties is 
not due to our coarse sampling of the S class and that splitting Ss in 
traditional stages does not improve the situation, our next step is to 
determine the appropriate binning of the S class that will bring together 
similar galaxies and put in different stages galaxies with different 
properties. This will give us a reason for the heterogeneity of the S class.
Optical (or ultraviolet) colors are a definite possibility. Using 
this parameter to discriminate between Ss of different ``stages", we find 
that bluer, and therefore strongly star forming, galaxies are much more 
segregated and have a larger velocity dispersion and higher brightness that 
redder spiral galaxies.  This means that the internal galaxy properties are 
correlated with the properties of the environment, immediately suggesting an 
environmental effect. Such an effect has already been detected as a (not 
statistically significant) difference in the luminosity function of S in high 
density vs low density regions (Dressler 1980b) and in clusters vs the field 
(Binggeli, Sandage \& Tammann 1988). 

With respect to early-type galaxies, whose space density (but not their 
properties) is determined by the environment, the situation is the opposite. 
The spiral spatial distribution is uniform in the field of Coma (this work) 
and in that of Perseus (Andreon 1993) and, therefore, the spiral space density is 
independent of the galaxy density (and of related quantities, such as distance 
from the cluster center and intergalactic gas density). All the internal 
properties of spirals measured with sufficient statistics show a trend with 
galaxy location.  We remind the reader that a uniform spiral distribution 
gives a spiral fraction rising with distance from the cluster center (or  
decreasing with density).  

The star formation induced by ram pressure naturally accounts for the
found segregation of blue spirals, for their higher mean surface brightnesses,
larger relative velocity with respect the cluster center, bluer color
in optical and in ultraviolet. The same mechanism has no effect on Es 
and S0s galaxies, since these galaxies are gas poor, naturally accounting 
for the fact that the internal properties of Es and S0s are environment 
independent.

\titleb {The re-interpretation of the observed morphological segregation}

The observed morphological segregation is the composition of two opposite 
effects: the dependence of the Es and S0s density on the whole galaxy 
density (and related quantities) and independence of the space density 
of Ss from the whole galaxy density. These two effects produce the known 
morphology-density or morphology-radius relations.  There is a large scatter
in both relations, so large that we did not detect them in Perseus and Coma. 

The fundamental parameter that determines the segregation is not the distance 
from the cluster center, as claimed by Whitmore, nor the galaxy density, as 
first claimed by Dressler, although the two exist, i.e. if you have enough 
statistics and you look for a dependence on one of these two parameters you 
find the segregation that you are looking for, as firstly shown by Sanrom\`a \&
Salvador-Sol\'e (1990). The fundamental parameter is 
not yet determined, but it is certainly related to the supercluster's main 
direction. In fact, early-type galaxies are segregated in a way related to the 
main direction of the supercluster in which they are embedded. 

The usual detection of a radial (or density) relation presented in
literature probably arises from the azimuthal average 
implicitly done by previous authors in 
studying the superposition of many different clusters
whose supercluster directions are (obviously) not aligned and/or 
in studying only
the radial dependency of the morphological segregation.

The stronger effect of the environment on galaxy properties,
the color dependence of spiral galaxies, i.e. their 
star forming rates, fails to be found by the standard morphological segregation 
which is related to changes in the galaxy morphological type. A consequence is 
that spiral colors, i.e. their star formation rate, is more sensitive to the 
environment than Hubble type, and it is this quantity that we have to 
investigate in order to study the effect of the environment on galaxy 
properties.

\titleb {The dichotomy of boEs and diE}

Starting with Bender et al. (1989), many authors have claimed 
there are differences in 
the properties (luminosity function, radio to optical luminosity, galaxy 
location with respect the cluster center ...) between small samples of boE 
with respect to even smaller samples of diEs (e.g. Longo et al. 1989, Shioya 
\& Taniguchi 1993, Caon \& Einasto 1995). An effort to enlarge the sample of 
galaxies, observing as many galaxies as possible {\it known} to be 
ellipticals, has been done by Nieto et al. (1991), Poulain, Nieto \& Davoust 
(1992), and by the ESO-Key program ``Toward a physical classification or 
early-type galaxies" (Bender et al. 1992). 

Because of the existence of ellipticals not catalogued as such or not 
catalogued at all, we follow another approach, more expensive from the 
observational point of view, with the aim of not missing any elliptical. We 
start to observe {\it complete} sample of galaxies, regardless of their 
morphological type. We think that the comparison of the results obtained on 
perhaps large but not fully controlled samples with those on smaller but well 
controlled samples could give some insights on the reality of the differences 
found (if any) between the properties of diE and boE. 

Complete samples fail to show differences between (bo+un)E and diE. In Coma, diEs 
are less spatially dispersed in the direction of the supercluster than other 
types, including (bo+un)E. Beside this, diEs and (bo+un)Es share the same 
properties (luminosity function, optical size, mean surface brightness, 
optical and infrared colors, radial, density and nearest galaxy 
distributions), within the large statistical errors due to the small 
statistics. In Perseus, (bo+un)E and diE share the same luminosity function and 
their spatial distribution presents a privileged direction, as in Coma. diE are 
more frequent than boE (+Epec) in two clusters studied with magnitude limited 
complete samples (14 vs 2 in Perseus and 15 vs 9 in Coma), but many galaxies 
are unclassified ellipticals (13 in Perseus and 11 in Coma), because of the 
larger distance of these two clusters compared to Virgo and the local field, 
and because of the existence of ellipticals which certainly are not diE and 
not boE. boE and diE in clusters share also the same radio-emission 
properties (Ledlow \& Owen 1995). 

On the contrary, in incomplete samples, often of the same size or smaller 
than our complete samples,
one finds differences between boE and diE. Bender et al. (1989) find that 
radio-loud galaxies are preferably boxy ellipticals,  Shioya \& Taniguchi 
(1993) find a larger frequency of boE in clusters, basing themselves on the 
incomplete and biased sample of Bender et al. (1989), rich in boxy radio-loud 
galaxies. Caon \& Einasto (1995) find that, in Virgo, boE are found in higher 
local galaxy density regions than diE. 

The two approaches thus lead to opposite conclusions on luminosity function, 
frequency, distance from the cluster center, local density and on the radio to 
optical luminosity of boE and diE. We think that the best conclusion we can 
draw is that differences, if any, between the properties of boE and diE are 
subtle and that the present samples are too small or too biased to enable one to 
detect them. 

\acknow{I thank my coworkers in the Coma project, E. Davoust and P. Poulain for 
permission to use our morphological types of the March 1995 observing run before 
publication. Furthermore, I wish to warmly thank E. Davoust for general advice and 
the carefull reading of the manuscript.
It is a pleasure to thank E. Salvador-Sol\'e, whose interest in 
this research pushed me to give the present emphasis to this work, and D. 
Maccagni and B. Garilli. 
I deeply thank J. Donas for providing me his ultraviolet catalogue of Coma
galaxies in advance of publication. This research would be not 
made possible without the generous allocation of telescope time (10 
observing runs) of the Comit\'e Fran\c{c}ais de Grand Telescopes, that I 
warmly tank. This research made use of the ``MAMA"  (Machine Automatique \`a 
Mesurer pour l'Astronomie) developed and operated by CNRS/INSU (Institut 
National des Sciences de l'Univers).} 

\begref{References}

\ref Andreon S., 1993, A\&A 276, L17

\ref Andreon S., 1994, A\&A 284, 801

\ref Andreon S., Davoust E., Nieto J.-L., Michard R., Poulain P., 1996, A\&AS, 
in press

\ref Beers T., Flynn K., Gebhardt K., 1990, AJ 100, 32

\ref Bender R., Capaccioli M., Macchetto F. et al., 1993, 
in Proceedings of the ESO/EIPC workshop ``Structure, dynamics and chemical 
evolution of elliptical galaxies", (ESO Conference and Workshop proceedings N 
45)Eds. I.J. Danziger, W.W. Zeilinger, K.Kj\"ar 

\ref Bender R., Surma P., D\"obereiner S. et al., 1989, A\&A 217, 35

\ref Binggeli B., Sandage A., Tammann G., 1988, ARA\&A  26, 509

\ref Bird C., 1994, AJ 107, 1637

\ref Biviano A., Durret F., Gerbal D. et al., 1995, A\&AS 111, 265

\ref Bothun G., Dressler A., 1986, ApJ 301, 57

\ref Boselli A., Gavazzi G., 1994, A\&A 283, 12

\ref Bower R., Lucey J., Ellis R., 1992, MNRAS 254, 601

\ref Bruzual G., Charlot S., 1993, ApJ 405, 538

\ref Briel U., Henry J., B\"ohringer H., 1992, A\&A 259, L31

\ref Bucknell M., Godwin J., Peach J., 1979, MNRAS 188, 579

\ref Butcher H., Oemler A., 1984, ApJ 285, 426

\ref Caldwell N., Rose J., Sharples R. et al., 1993, AJ 106, 473

\ref Caon N., Einasto M., 1995, MNRAS 273, 913

\ref Colless M., Dunn A., 1996, ApJ, 458, 435

\ref Davis M., Geller M., 1976, ApJ 208, 13

\ref Donas J., Milliard B., Laget M., 1995, A\&A,  303, 661

\ref Dressler A., 1980a, ApJS 42, 565

\ref Dressler A., 1980b, ApJ 236, 351

\ref Dressler A., Oemler A., Sparks W. et al. 1994, ApJ 435, L23

\ref Escalera E., Biviano A., Girardi M. et al., 1994, ApJ 423, 539

\ref Fitchett M, Webster R., 1987 ApJ 317, 653 

\ref Garilli B., Bottini D., Maccagni D. et al. 1996, ApJS, in press

\ref Gavazzi G., 1987, ApJ 320, 96

\ref Gavazzi G., Trinchieri G., 1989, ApJ 342, 718

\ref Godwin J., Metcalfe N., Peach J., 1983, MNRAS 202, 113

\ref Godwin J., Peach J., 1977, MNRAS 181, 323

\ref Gunn J., Gott R., 1972, ApJ 170, 1

\ref Guzm\'an R., Lucey J., Carter D., Terlevich R.,  1992, MNRAS 257, 187

\ref Holmberg E., 1958, Medd. Lund. Astr. Obs. Ser. II, No. 136

\ref Hubble E., 1926, ApJ 64, 321

\ref Hubble E., Humason M., 1931, ApJ 74, 43

\ref Jones C., Forman W., 1984, ApJ 276, 38 

\ref J\o rgensen I., Franx M., Kjaergaard P., 1993, ApJ 411, 34

\ref J\o rgensen I., Franx M., Kjaergaard P., 1996, MNRAS, in press

\ref Huchra J. 1985, The Virgo Cluster, ESO Workshop Proceeding No. 20, eds Richter 0., Binggeli B., (Munich: ESO), p. 181

\ref Kent S., Gunn J., 1982, AJ 87, 945

\ref Larson R., Tinsley B., Caldwell N., 1980, ApJ 237, 692

\ref Ledlow M., Owen F., 1995, AJ 110, 1959

\ref Lonsdale C., Persson E., Matthews K., 1984, ApJ 287, 95

\ref Longo G., Capaccioli M., Bender R. et al., 1989, A\&A 225, L17

\ref Malumuth E., Kriss G., Van Dyke Dixon W. et al., 1992, AJ 104, 495

\ref Mellier Y., Mathez G., Mazure A. et al. 1988, A\&A 199, 67

\ref Nieto J.-L., Bender R., Poulain P. et al. 1992, A\&A 257, 97

\ref Nieto J.-L., Poulain P., Davoust E., 1994, A\&A 283, 1

\ref Nieto J.-L., Poulain P., Davoust E. et al. 1991, A\&AS 88, 559

\ref Pahre M., Djorgovski S., de Carvalho R., 1996, ApJL, in press

\ref Postman M, Geller M., 1984, ApJ 281, 95

\ref Poulain P., Nieto J.-L., Davoust E., 1991, A\&AS 95, 129

\ref Recillas-Cruz E., Carrasco L., Serrano A. et al., 1990, 
A\&A 229, 64

\ref Roberts M., Haynes M., 1994, ARA\&A 32, 115

\ref Saglia R., Bender R., Dressler A., 1993, A\&A 179, 77

\ref Salvador-Sol\'e E., Gonz\'alez-Casado G., Solanes J., 1993, ApJ 410,1

\ref Salvador-Sol\'e E., Sanrom\`a M., Gonz\'alez-Casado G., 1993, ApJ 402, 398

\ref Sandage A., Binggeli B., Tammann G., 1985, AJ 90, 1759

\ref Sanrom\`a M., Salvador-Sol\'e E., 1990, ApJ 360, 16

\ref Sarazin C., 1986, {\it X-ray emissions from clusters of galaxies}, Cambridge
University Press

\ref Schneider D., Dressler A., Gunn J., 1986, AJ 92, 523

\ref Shioya Y., Taniguchi Y., 1993, PASJ 45, L39

\ref Staveley-Smith L, Davies R., 1988, MNRAS 224, 953

\ref Staveley-Smith L, Davies R., 1988, MNRAS 231, 833

\ref Thompson L., Gregory S., 1980, ApJ 242, 1

\ref de Vaucouleurs G., de Vaucouleurs A., Corwin H et al. 1991 (New York: Springer Verlag)

\ref Visvanathan N., Sandage A., 1977, ApJ 216, 214 

\ref Whitmore B., 1984, ApJ 278, 61

\ref Whitmore B., Gilmore D., Jones C., 1993, ApJ 407,489

\ref White S., Briel U., Henry J., 1993, NMRAS 261, L8

\ref Zabludoff A., Franx M., 1993, AJ 106, 1314

\endref

\vfill \eject
\null
\vfill \eject

\begtabfullwid
\tabcap{1}{Probability, expressed in percentage, that two galaxy classes 
have the same parental luminosity, optical size, mean
surface brightness, B-R color, infrared colors, x, y, radial, azimuthal,
distance from the nearest galaxy, density and velocity distribution (or,
better, one minus the confidence level to reject the hypothesis that
the types are drawn from the same parental distribution). Only
pairs of morphological types having less than 0.2 \% of probability to be
drawn from the same parental distribution are
listed. For the sake of clarity, probabilities larger than 15 \% are replaced
with blanks. Null detections are not listed but are discussed in the text
when interesting.} 
{\small
\def\tvi{\vrule depth 5pt width 0pt}
\def\tv{\tvi\vrule}
\offinterlineskip
\halign{#\hfill&\ #\hfill&\ #\hfill&\ \hfill#&\ \hfill#&\ \hfill#&\ \hfill#&\ \hfill#&\ \hfill# \ &\tv#& \quad #\hfill&\ #\hfill&\ #\hfill&\ \hfill#&\ \hfill#&\ \hfill#&\ \hfill#&\ \hfill#&\ \hfill#\cr
\noalign{\hrule\medskip}
       c1&  c2&      quant.&    P5\hfill& T\hfill& F\hfill&  SI\hfill& TI\hfill&  KS\hfill& &      c1&  c2&     quant.&    P5\hfill& T\hfill& F\hfill&  SI\hfill& TI\hfill&  KS\hfill\cr
\noalign{\medskip\hrule\medskip}
%
&&&&&&&&&&&&&&&\cr
\multispan9{\hfill {\it \underbar{ Detailed morphological classes }} \hfill}&&\multispan9{\hfill {\it \underbar{ Hubble morphological classes }} \hfill} \cr          
 (un+bo)E& SA0&         B&   0.000&  0.000&  1.954&       &       &   0.005&&        E&  S0&         B&   0.000&  0.000&  0.436&       &       &   0.009\cr
 (un+bo)E& SB0&         B&   0.001&  0.058&       &       &       &   0.771&&       E&   S&         B&   0.006&  0.058& 13.201&       &       &   0.090\cr
 (un+bo)E&   S&         B&   0.009&  0.054&       &       &       &   0.103&&        E&  S0& $r_{iso}$&   0.000&  0.000&  0.046&       &       &   0.000\cr
      diE& SA0&         B&   0.074&  1.099&  9.568&       &       &  13.369&&        E&   S& $r_{iso}$&   0.000&  0.000&  0.012&       &       &   0.000\cr
      SA0&   S&         B&   0.142&       &  3.217&       &       &        &&        E&  S0&  $<$SuBr$>$&   0.000&  0.008&       &       &  4.958&   0.223\cr
 (un+bo)E& SA0& $r_{iso}$&   0.000&  0.000&  0.519&       &       &   0.000&&        E&   S&  $<$SuBr$>$&   0.001&  0.012&  5.426&       &       &   0.022\cr
 (un+bo)E& SB0& $r_{iso}$&   0.001&  0.009& 14.544&       &       &   0.013&&       S0&   S&  $<$SuBr$>$&   0.156&       & 12.748&       & 10.285&  10.141\cr
 (un+bo)E&   S& $r_{iso}$&   0.000&  0.000&  0.851&       &       &   0.000&&        E&   S&     $B-R$&   0.000&  0.135&  0.168&  0.072&       &   0.004\cr
      diE& SA0& $r_{iso}$&   0.061&  0.775&  5.366&       & 12.380&  11.590&&       S0&   S&     $B-R$&   0.000&  0.021&  0.025&  0.729&       &   0.004\cr
      diE&   S& $r_{iso}$&   0.034&  0.474&  2.644&       &       &   7.233&&         E&   S&     $H-K$&   0.000&  0.072&  0.669&       & 12.444&   0.805\cr                                                                     
 (un+bo)E& SA0& $<$SuBr$>$&   0.005&  0.015&       &       &       &   0.125&&       S0&   S&     $H-K$&   0.108&  2.754&  9.526&       &       &   4.756\cr
 (un+bo)E& SB0& $<$SuBr$>$&   0.164&  0.597&       &       &       &   2.350&&        E&   S&         x&   0.115&       &  2.366&  2.908& 14.701&        \cr
 (un+bo)E&   S& $<$SuBr$>$&   0.000&  0.007&  9.953&       &       &   0.031&&      S0&   S&         y&   0.017&       &  2.429&       &  0.130&  11.636\cr
 (un+bo)E& SB0&     $B-R$&   0.076&       &  5.288&  1.856&       &        &&        E&   S&     $d_1$&   0.005&  0.203&       &       &       &   0.054\cr
 (un+bo)E&   S&     $B-R$&   0.000&  2.122&  1.085&  0.274&       &   0.067&&       E&   S&         v&   0.059&       &  1.200&       &       &  12.973\cr
      diE&   S&     $B-R$&   0.000& 12.344&  2.857&  0.001&       &   0.059&&       S0&   S&         v&   0.148&       &  3.206&       &       &        \cr  
      SA0&   S&     $B-R$&   0.000&  1.533&  0.270&  0.042&       &   0.028&&\multispan9{\quad\hrulefill}\cr 
      SB0&   S&     $B-R$&   0.076& 13.227&  0.453&       &       &   0.376&&\cr
 (un+bo)E& SB0&     $H-K$&   0.083&  0.334&       &       &       &   1.943&&\multispan9{\hfill {\it \underbar{ Color classes }} \hfill}\cr
 (un+bo)E&   S&     $H-K$&   0.001&  0.272&  0.584&       &       &   0.232&&  red& blue&   $<$SuBr$>$&   0.000&  0.000&       &       &       &   0.008\cr 
 (un+bo)E& diE&         x&   0.030&  7.205&  1.376&       &       &  13.178&&  red& blue&            y&   0.125&       & 13.679&       &  3.894&  11.641\cr
      diE& SA0&         x&   0.062&       &  0.703&       &       &  11.985&&  red& blue&        $d_1$&   0.000&  0.119&       &       &       &   0.002\cr
      diE& SB0&         x&   0.141&       &  1.701&       &       &        &&  red& blue&  $\rho_{10}$&   0.005&  0.720&  1.981&  3.401&       &   0.046\cr
      diE&   S&         x&   0.125&       &  0.327&       &       &        &&  red& blue&            r&   0.039&  2.496&  0.948&       &       &   0.112\cr
      diE& SB0&         y&   0.086& 10.451& 13.855&       &       &        &&  red& blue&            v&   0.113&  3.913&  5.782&       &       &   2.887\cr
      SA0& SB0&         y&   0.024&       &  0.850&  9.062& 10.609&        &&\multispan9{\quad\hrulefill}\cr    
      SB0&   S&         y&   0.000&       &  0.057&  2.679&  0.548&   5.855&&\cr                                                   
 (un+bo)E&   S&     $d_1$&   0.097&  1.713&       &       &       &   0.799&&\multispan9{\hfill {\it \underbar{ Red and blue spirals }} \hfill}\cr
      diE&   S&     $d_1$&   0.046&  0.396& 13.983&  8.026&       &   0.365&&  red S& blue S& $\rho_{10}$& 0.024&  0.213& 7.256&       &       &   0.694\cr
      diE&   S&         r&   0.147&  0.813&  6.752&       &       &   1.617&&  red S& blue S&  $<$SuBr$>$& 0.044&  0.234&      &       &       &   1.161\cr
 (un+bo)E& SB0&         v&   0.040&  6.837& 13.436& 10.661&       &        &&  red S& blue S&        UV-B& 0.000&  0.121&      &       &       &   0.027\cr
 (un+bo)E&   S&         v&   0.013&       &  0.145&       &       &   8.755&&\cr
      SA0& SB0&         v&   0.174&  3.951&       &  6.576&       &        &&\cr                                                                            
      SA0&   S&         v&   0.121&       &  1.317&       &       &   5.192&&\cr
\noalign{\medskip\hrule\medskip}}

} 
\endtab

\null
\vfill \eject

\null
\vfill \eject

\psfig{figure=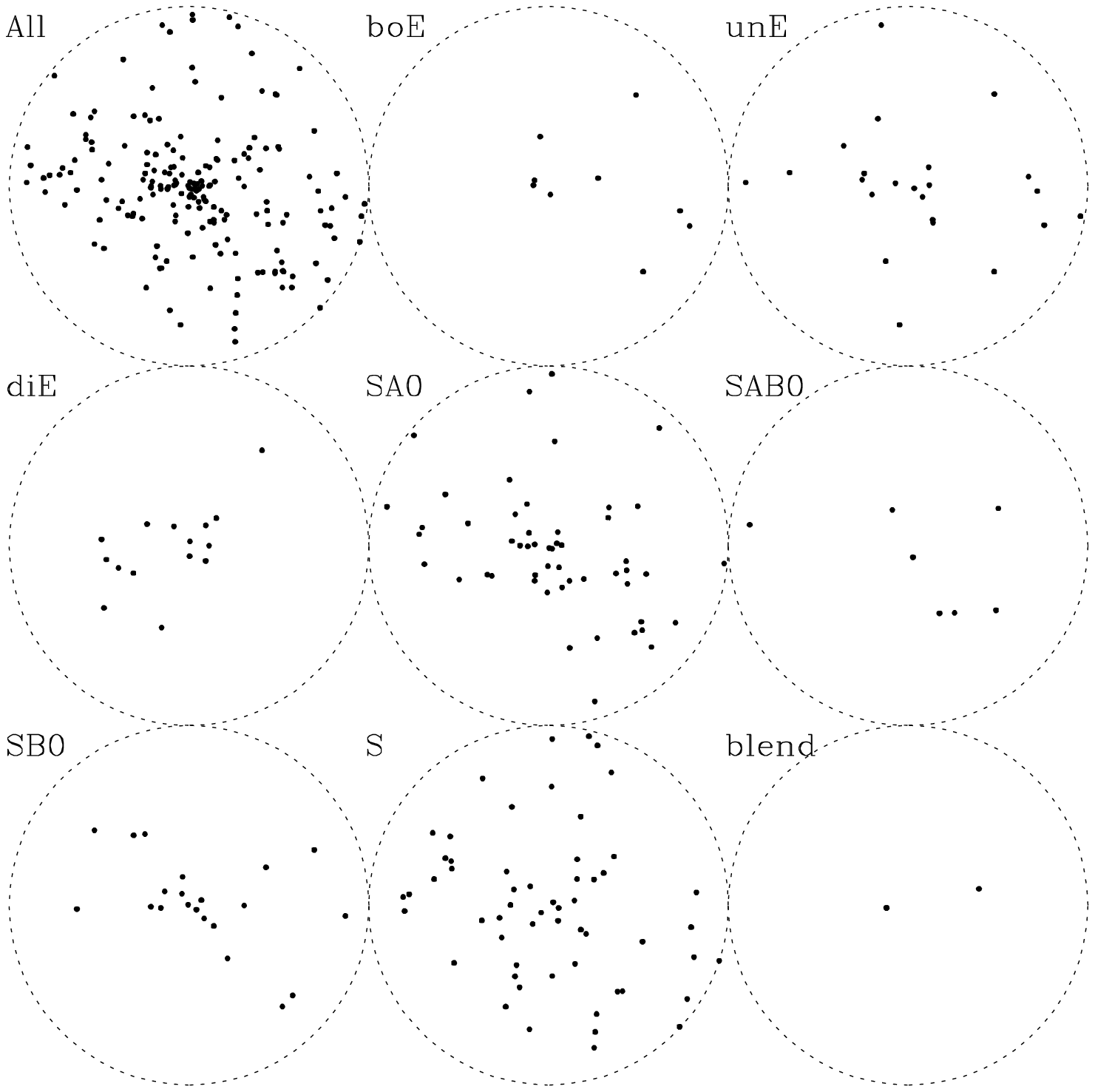,width=8.5truecm,bbllx=20mm,bblly=75mm,bburx=175mm,bbury=235mm}
\figure{1}{Spatial distribution of all galaxies of Coma and of each individual Hubble
type. North is up and East is left. The x,y coordinate system used in this work 
is rotated 45 degrees clockwise to align it with the Coma/A1367 supercluster. 
Note that there are 187 points (galaxies) in the top-left panel. In this 
small plot some close points are superposed and thus undistinguishable on the 
diagram.} 

\psfig{figure=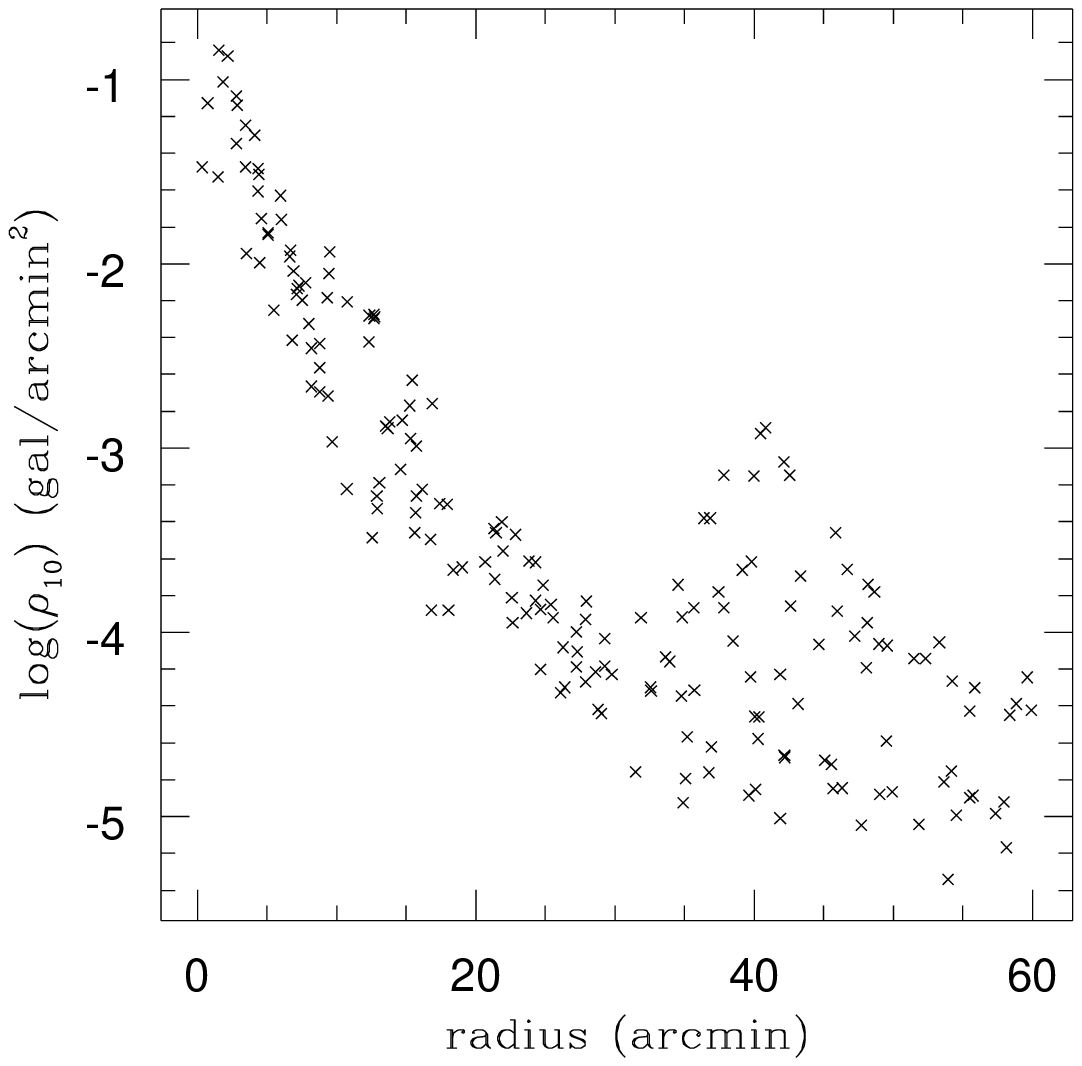,width=8.5truecm,bbllx=20mm,bblly=65mm,bburx=135mm,bbury=180mm}
\figure{2}{Relation between the clustercentric distance and the local
density in Coma.  Each point corresponds to one galaxy.}

\null
\vfill \eject
\null
\vfill \eject

\hbox{
\psfig{figure=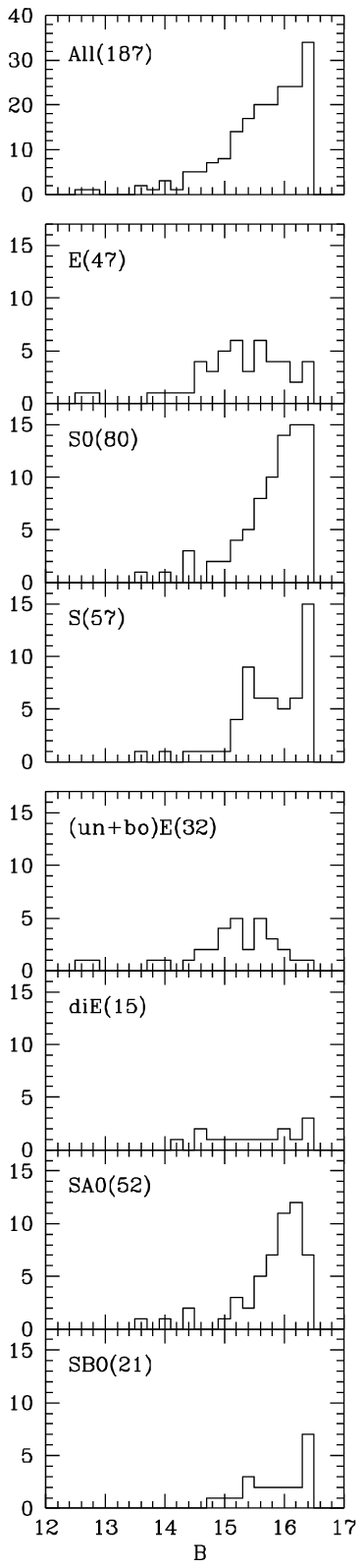,width=4.25truecm,height=21truecm,bbllx=52mm,bblly=80mm,bburx=92mm,bbury=245mm} \break
\psfig{figure=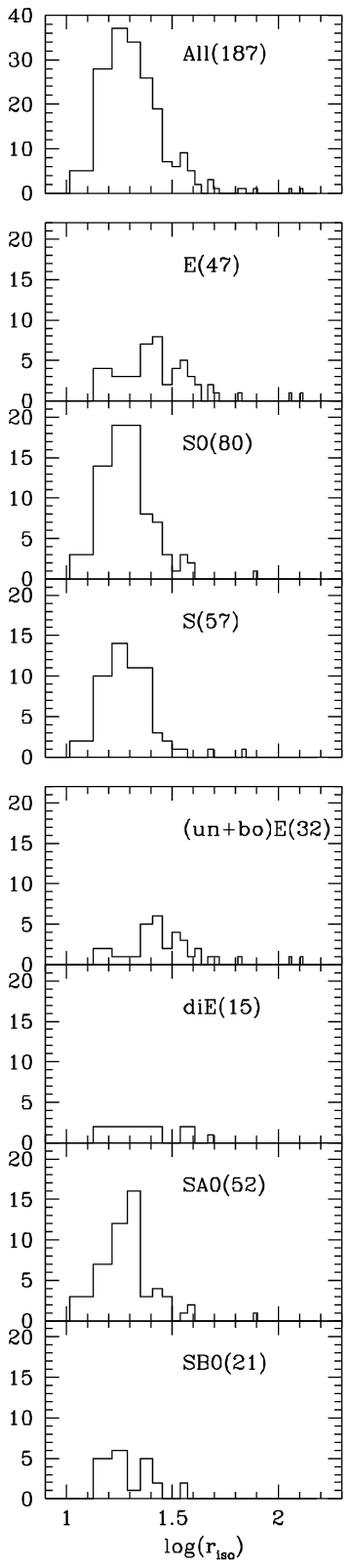,width=4.25truecm,height=21truecm,bbllx=52mm,bblly=80mm,bburx=92mm,bbury=245mm} \break
\psfig{figure=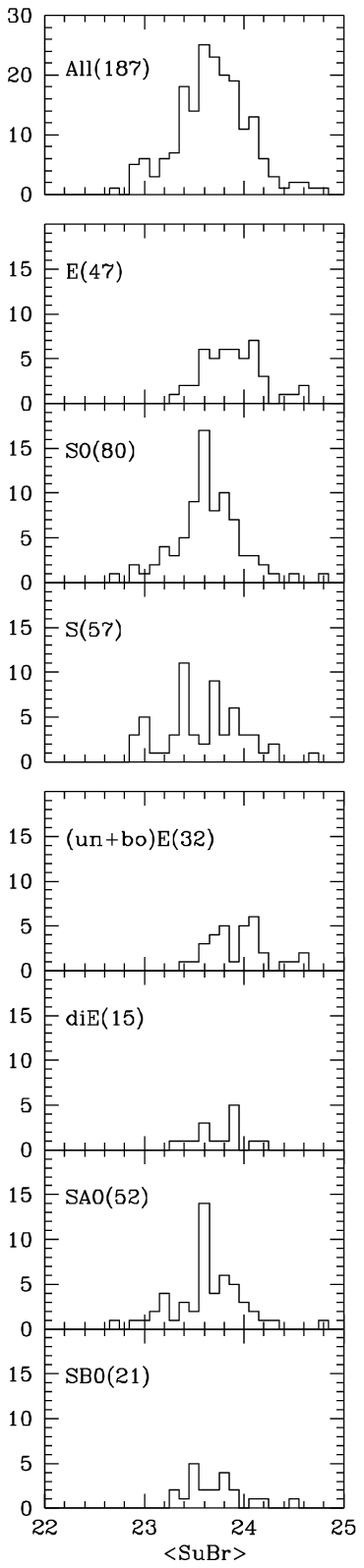,width=4.25truecm,height=21truecm,bbllx=52mm,bblly=80mm,bburx=92mm,bbury=245mm} \break
\psfig{figure=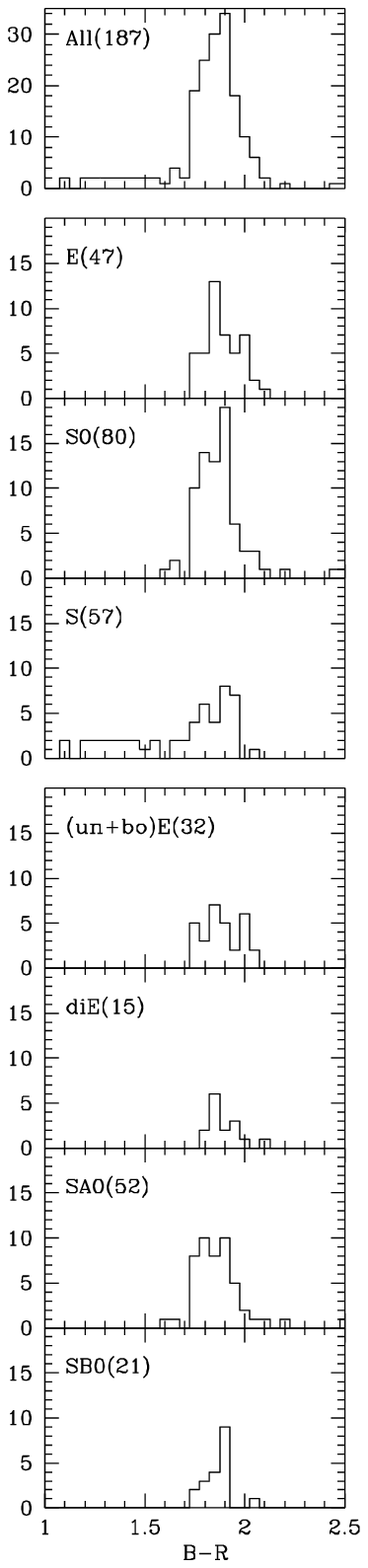,width=4.25truecm,height=21truecm,bbllx=52mm,bblly=80mm,bburx=92mm,bbury=245mm} \break
}
\figure{3a}{Luminosity function, size, mean surface brightness and B-R color distributions
of all galaxies of Coma and of each individual Hubble type. Radii are measured in arcsec and 
surface brightnesses are measured in mag arcsec$^{-2}$}

\null
\vfill \eject
\null
\vfill \eject

\hbox{
\psfig{figure=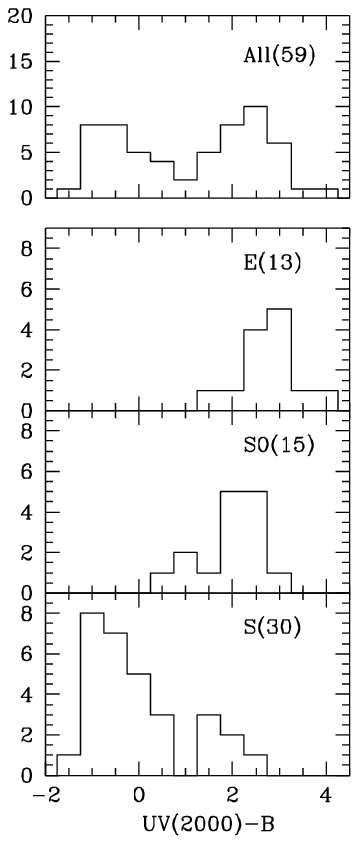,width=4.25truecm,height=21truecm,bbllx=52mm,bblly=80mm,bburx=92mm,bbury=245mm} \break
\psfig{figure=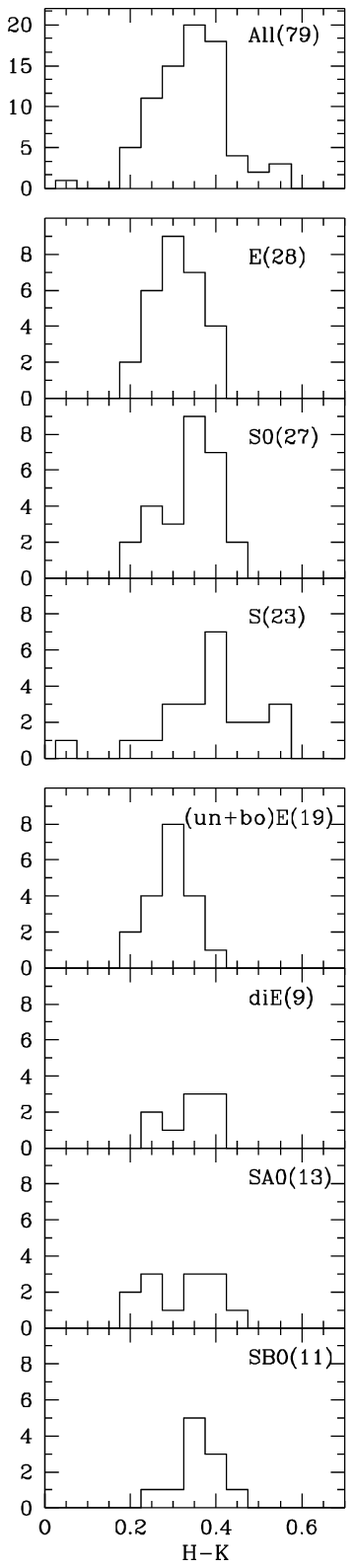,width=4.25truecm,height=21truecm,bbllx=52mm,bblly=80mm,bburx=92mm,bbury=245mm} \break
\psfig{figure=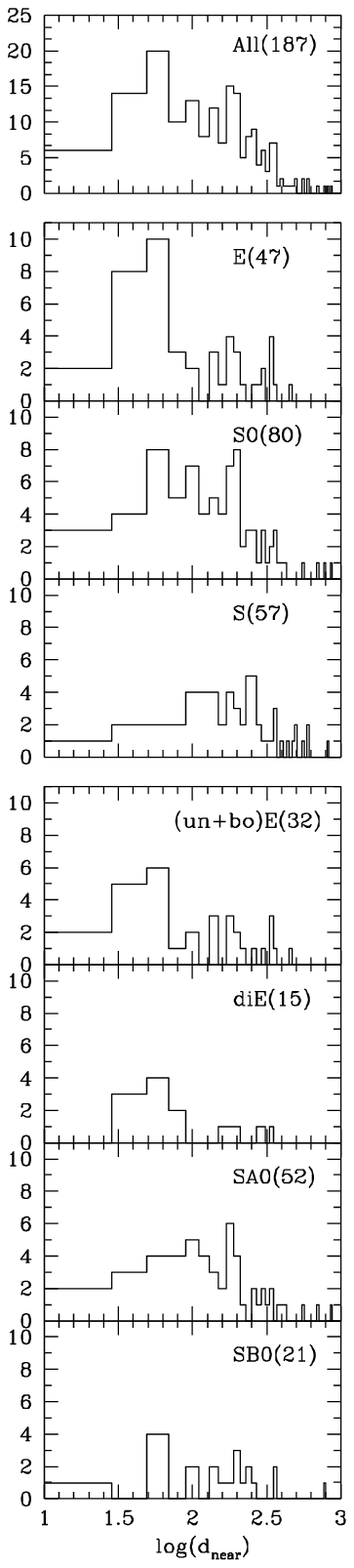,width=4.25truecm,height=21truecm,bbllx=52mm,bblly=80mm,bburx=92mm,bbury=245mm} \break
\psfig{figure=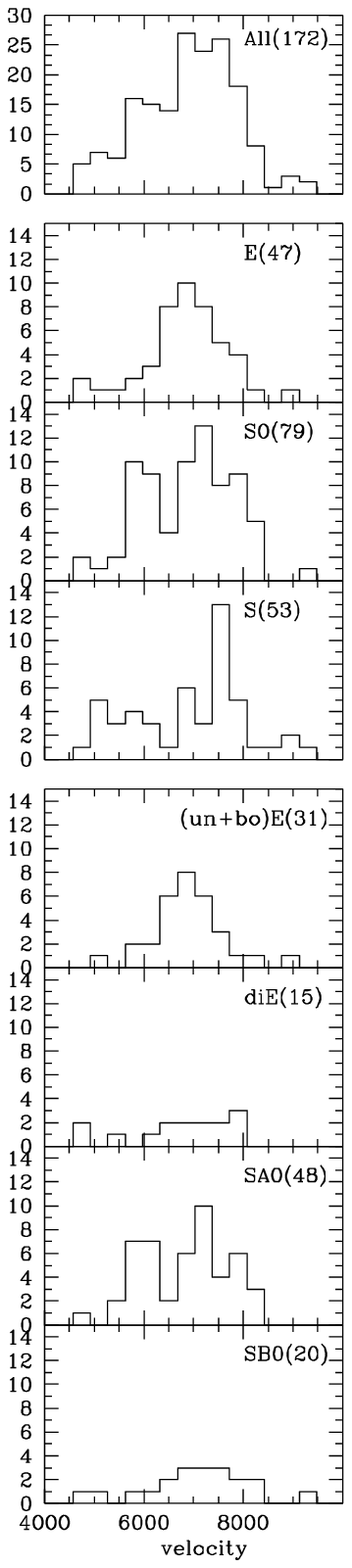,width=4.25truecm,height=21truecm,bbllx=52mm,bblly=80mm,bburx=92mm,bbury=245mm} \break
}
\figure{3b}{Ultraviolet (UV(2000)-B) and infrared $H-K$ color, distance from 
the nearest galaxy and velocity distributions of all galaxies of Coma and of 
each Hubble type. Please note that no correction was applied to raw $H-K$ 
color. Distances are measured in arcsec and velocities are measured in km 
s$^{-1}$. Due to small statistics, the ultraviolet color is shown only for the 
coarse classes.}

\null
\vfill \eject
\null
\vfill \eject

\psfig{figure=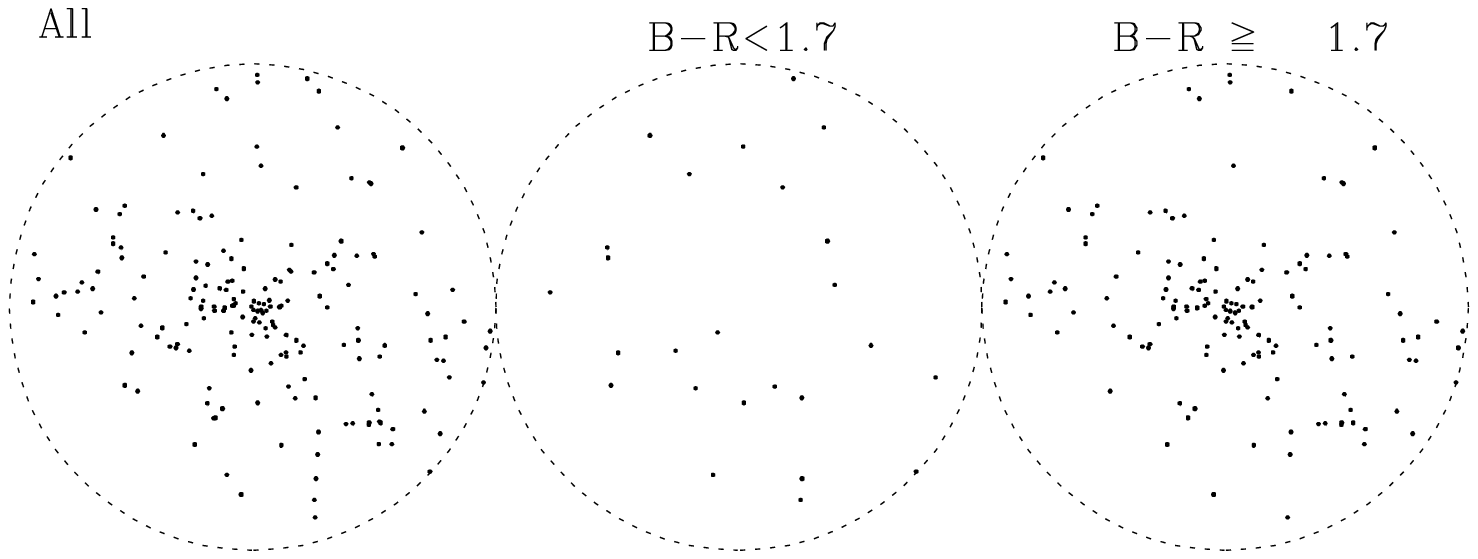,width=8.5truecm,bbllx=25mm,bblly=185mm,bburx=175mm,bbury=245mm}
\figure{4}{Spatial distributions of all galaxies of Coma and of red ($B-R
\ge 1.7$) and blue ($B-R<1.7$) galaxies}

\psfig{figure=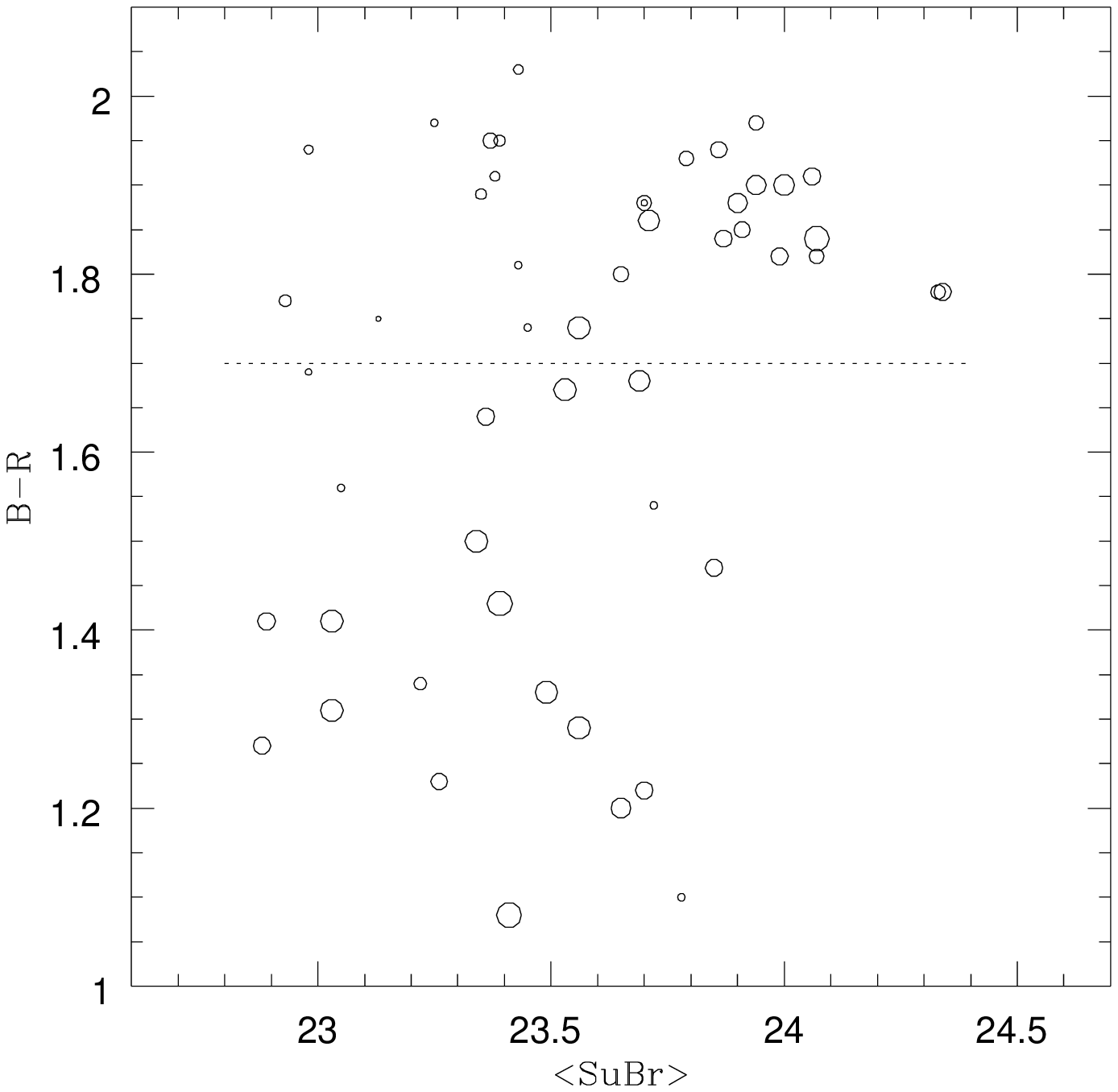,width=8.5truecm,bbllx=20mm,bblly=70mm,bburx=165mm,bbury=210mm}
\figure{5}{Optical color vs. mean surface brightness for spirals in Coma (57 
galaxies) The dashed line indicates our division of the sample into blue ($B-
R<1.7$) and red spirals. The size of the symbols is proportional to the galaxy 
ellipticity. Round galaxies are therefore small circles.}

\bye